\documentclass[aps,prb,twocolumn,amsmath]{revtex4} 

\usepackage{dcolumn}
\usepackage{bm}
\usepackage{color,graphicx}
\usepackage{amsfonts}
\usepackage{amssymb}
\usepackage{rotate}

\usepackage{subfigure}
\usepackage[latin1]{inputenc}
\usepackage{subfigure}
\usepackage{amsmath}
\usepackage{dcolumn}

\renewcommand{\-}{\,-\,}

\let\oldmarginpar\marginpar
\renewcommand\marginpar[1]{\-\oldmarginpar[\raggedleft\tiny #1]%
{\raggedright\tiny #1}}

\begin{document}

\title{\textbf{Magnetic multipole analysis of kagom\'e and artificial ice dipolar
    arrays}} 

\author{Gunnar M\"{o}ller$^1$ and Roderich Moessner$^2$}
\affiliation{${}^1$TCM Group, Cavendish Laboratory, J.~J.~Thomson Ave.,
  Cambridge CB3 0HE, UK\\ 
${}^2$Max Planck Institute for the Physics of Complex Systems, Dresden, Germany}

\date{June 25, 2009}

\begin{abstract}
We analyse an array of linearly extended monodomain dipoles forming square and
kagom\'e lattices. We find that its phase diagram contains two (distinct)
finite-entropy kagom\'e ice regimes---one disordered, one algebraic---as well
as a low-temperature ordered phase. In the limit of the islands almost
touching, we find a staircase of corresponding entropy plateaux, which is
analytically captured by a theory based on magnetic charges. For the case of a
modified square ice array, we show that the charges (`monopoles') are
excitations experiencing two distinct Coulomb interactions: a magnetic
`three-dimensional' one as well as a logarithmic `two dimensional' one of
entropic origin.  
\end{abstract}

\maketitle

An intensifying experimental effort is under way aimed at constructing dipolar
arrays of monodomain nanomagnets.\cite{Wang,Cummings,Zabel,Tanaka} From a
technological viewpoint, this is a promising approach as it harnesses mature
techniques---for example those involved in the lithographic preparation of
arrays---often highly developed in an industrial context. Indeed, one of the
fundamental questions from this perspective is to understand interactions and
dynamics of closely packed nanomagnets as these are among the factors
determining performance limits on high-density magnetic storage media.  

From a many-body physics angle, this forms one aspect of a wider program
of studying artificial systems which can be constructed, manipulated and probed
on the level of an individual degree of freedom. 
Pioneering experiments by Wang \emph{et al.} on the two-dimensional `square ice' array 
have established a proof-of-principle of the feasibility of such an enterprise\cite{Wang} 
by demonstrating that the magnets assume configurations reflecting
their interactions, paving the way to a study of their collective
properties.\cite{Wang,MollerMoessner}

While the study of dipolar magnets is a rich subject with a venerable tradition\cite{Luttinger-Tisza} 
with a recently rejuvenated activity in the context of cold
gases\cite{MenottiPfau}---interesting on account of the long ranged
and anisotropic nature of the interactions---recent work on the spin ice
compounds\cite{BramwellGingras} has found a number of surprising new
phenomena. In particular, it was found that even the complex dipolar
interactions can lead to large low-temperature degeneracies, apparently with
little additional fine-tuning.\cite{RamirezNat,SiddShas,GingrasEnt} A simple
model which considered each magnetic dipole as a pair of nearly opposite
magnetic charges (``monopoles'') was able to account for this phenomenon
analytically, and established that in spin ice, these monopoles are the
appropriate degrees of freedom at low temperatures.\cite{CastelNature}

In this paper, we analyse interactions in artificial dipolar arrays in $d =
2$. We study two geometries, the qualitative behavior of which varies
reflecting simple lattice parameters such as even/oddness of
their coordination. Besides the abovementioned artificial spin ice, we
particularly analyse the case where the dipolar islands reside on the bonds of
a honeycomb lattice with their centres forming a kagom\'e array, which has been a
focus of recent experimental studies.\cite{Tanaka,Cummings,Zabel,mengotti} For this kagom\'e
array, we find a series of regimes: firstly, the paramagnet corresponds to a
gas of charges $\pm$1 and $\pm$3. As the interaction is increased in strength,
there is a crossover to a gas of charges $\pm$1 (kagom\'e ice I, K1) which
terminates in an ordering transition of the charges in a two-sublattice NaCl
structure on the honeycomb lattice (kagom\'e ice II, K2). There being
exponentially many configurations of the islands which yield a NaCl charge
configuration  a final transition into an ordered structure with tripled unit
cell removes this entropy. In a particular limit, we can extend the monopole
theory for spin ice to a multipole one which captures all these regimes in analytic detail.

We supplement this by an analysis of the square ice array, where we find a
hybrid Coulomb interaction between monopoles: one is the usual 1/r interaction between
magnetic charges, whereas the other is of entropic origin and exhibits a
logarithmic distance dependence. An additional linear confining force can be
removed by shifting the monopoles in the third dimension so that
their locations define tetrahedra. In closing, we remark on connections to
other systems as well as issues of dynamics and finite-size effects.

\noindent {\bf Dipolar needles:}~ The basic degree of freedom is a dipolar
island which we model with the 
following (idealised) properties. The magnetic moment density $\vec\mu/l$ is
uniform along the length, $l$, of the island. The islands are monodomain with
the moment pointing along their axis, yielding an effective Ising spin---as
was verified experimentally.\cite{Lederman,ChouIEEE,Cummings,Wang,Zabel,mengotti} We take
the shape of the island to be that of a needle, with a 
length much greater than its width, neglecting its finite transverse extent. A
needle with uniform magnetisation density $q=|\vec\mu|/l$ sets up the same
potential $U(\vec{R})$ as that of a pair of magnetic charges $\pm q$ located at
its tips (in the context of micromagnetic domain walls, 
see Refs.~\onlinecite{Saitoh,Tchernyshyov}). The value of the Ising spin encodes which of the
ends---located at $\vec{r}_a$ and $\vec{r}_b$---hosts the positive charge:  
\begin{equation}
U(\vec{R})=\frac{\mu_0\mu}{4\pi l}\int d{\vec{l}}\cdot
\vec{\nabla} \frac{1}{|\vec{r}-\vec{R}|} = \frac{\mu_0 q}{4 \pi}\left[
  \frac{1}{|\vec{r}_a-\vec{R}|} - \frac{1}{|\vec{r}_b-\vec{R}|}\right]  . 
\nonumber
\end{equation}

In the following, we express lengths in units of $a$, the nearest-neighbor
distance of the respective lattices. The dipolar islands are located on the
bonds of the lattice. In the regime $l/a\to 1$, which is approximately realized
in recent experiments,\cite{Tanaka,Cummings,Zabel} it is convenient to define
$\epsilon=1-l/a$. We generally use $q^2/a$ as the unit of energy, but shall
also refer to the nearest neighbor Ising coupling strength of the islands, $J$,
where this is more instructive. 

Many of the salient features of these spin systems can be deduced intuitively
in the charge picture which, in the limit of $\epsilon \to 0$, admits an
analytic treatment.

Let us first consider the case of the kagom\'e array. The vertices of the kagom\'e
lattice can have total charge $Q=\pm 1$ (one-out, two-in or 
two-out, one-in) or $Q=\pm 3$ (all in or all out). The corresponding
on-site energies of $E_1 = \frac{-2}{\sqrt{3}\epsilon a}$ and $E_3 =
\frac{2\sqrt{3}}{\epsilon a}$ are such that the latter appear only at high
temperature, $T$, in a phase that can be classified as a paramagnet. Below
$T_\text{ice}^I\sim E_3-E_1 \sim 2J$, the population of charge
$\pm 3$ vertices is exponentially suppressed. This crossover occurs without
change of symmetry. The remaining configurations with $Q=\pm 1$ everywhere
are precisely those satisfying the ice-rule (two-in, one-out or vice
versa). The tendency to this kind of ``order'' was investigated and
confirmed in recent
experiments.\cite{Tanaka,Cummings,Zabel}

Unlike for the case of spin-ice, where a coordination number of $4$ allows for
vertex charges to vanish (and finite charges are excitations in $d = 3$, see
Ref.~\onlinecite{CastelNature}), the odd coordination of $3$ enforces the presence of
magnetic charges in kagom\'e ice at any temperature.  

This kagom\'e ice (K1) regime is a highly frustrated phase of matter, with an
entropy (known exactly) about $5/7$ that of a free spin,
$\mathcal{S}_\text{ice}^{I} \approx \frac{2}{3}\ln\frac{3}{\sqrt{2}} \approx 0.501$.

At lower temperatures $T_\text{ice}^{II}\sim q^2/a$, interactions between
vertices assert themselves. They lead to a ``charge-ordered'' NaCl ionic
crystal, in which each sublattice of the honeycomb lattice only hosts
charges $|Q|=1$ of a given sign---the resulting magnetic charge order is thus
described by an Ising order parameter. 

\begin{figure}[ttbb]
  \begin{center}
    \includegraphics[width=0.95\columnwidth]{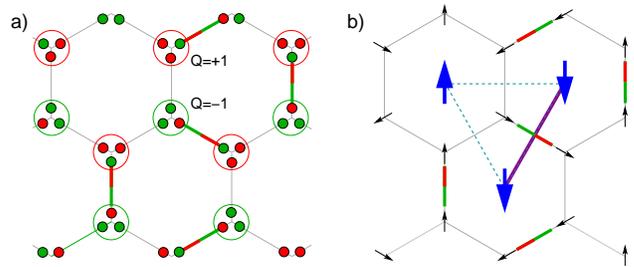}\\
    \caption{(color online) a) Orientations of Ising spins on the links are mapped to a
      distribution of magnetic charges located at the extremities of
      each island. In kagom\'e-ice, the total charge at each vertex is
      constrained to $|Q|=1$. At low temperatures, vertex charges order such
      that all vertices of sublattice $\alpha$ have charge
      $Q_\alpha=(-1)^\alpha$ (as shown). To count the configurations
      corresponding to that NaCl charge-state, we map these to dimer coverings of
      the honeycomb lattice, where oriented dimers are present at all links
      connecting the minority charges of the adjacent vertices. 
      b) To count the dimer coverings of the hexagonal lattice, we note that
      dimers present on the hexagonal lattice can be mapped to broken links of
      the antiferromagnet on the dual triangular lattice. The particular
      configuration shown contains the unit-cell of the groundstate, formed by
      the three highlighted plaquettes. Arrows shown at the vertices display
      the direction of the local dipole moment.} 
    \label{fig:mappings}
  \end{center}
\end{figure}

While the NaCl structure is fully ordered in terms of charges, it is still
consistent with an exponentially large number of configurations of the
underlying Ising degrees of freedom of the islands. This kagom\'e ice II (K2)
regime is therefore only partially ordered. To calculate this entropy we map
each spin configuration in kagom\'e ice II to a dimer covering of the hexagonal
lattice: note that at any vertex, there is precisely one minority magnetic
charge. As $Q_\alpha=-Q_\beta$, the edge with the 
minority charge is necessarily connected to the minority charge at the
neighboring vertex, as shown in Fig.~\ref{fig:mappings}a). Thus, each charge
configuration singles out an (oriented) dimer covering, with precisely one
dimer pointing from the negative to the positive charge impinging on each
vertex. The orientation of the dimers encodes the broken Ising symmetry.

The entropy of hexagonal dimer coverings is known by a mapping to the triangular
Ising antiferromagnet with a well-known groundstate entropy 
$\mathcal{S}_\text{AFM}^\triangle$.  Adjusting for the number of elementary
degrees of freedom---islands rather then dimers---we find an entropy per island
of 
\begin{equation}
\mathcal{S}_\text{ice}^{II} = \frac{1}{3}\mathcal{S}_\text{AFM}^\triangle
\approx 0.108\,.
\end{equation}

Finally, at very low temperatures, further neighbor interactions induce a
transition into an ordered phase which removes all remaining entropy. As the
most favourable arrangement of a single hexagonal plaquette is for all islands
to point head-to-tail,\cite{mengotti} it is natural to guess that the state
should optimise 
the occurrence of that motif. However, no full tiling of the honeycomb lattice
with such oriented loops is possible. Instead, one finds a configuration, in
which 2/3 of all plaquettes are in such a state, which is precisely the one
shown in  Fig.~\ref{fig:mappings}. The three plaquettes highlighted in
Fig.~\ref{fig:mappings}b) form the tripled $\sqrt{3} \times \sqrt{3}$ unit-cell
of the crystal with periodicity in the lattice vectors $v_1=(0,3a)$ and
$v_2=(3\sqrt{3}a/2,3a/2)$. We will refer to this groundstate as the `loop state', below.
Note that the plaquettes without an oriented loop host no dimers at all.

\noindent {\bf Analytic magnetic multiple expansion:}~The cascade of regimes
outlined here can analytically be accounted for in the limit $\epsilon \to 0$,
as each transition/crossover is associated with a parametrically 
different energy scale. This approach amounts
to a development of a multipole expansion beyond the dumbbell model of spin
ice. One finds for the energy of the loop state (expressed per hexagonal site):
\begin{equation}
\label{eq:LoopEnergy}
E = \frac{q^2}{a}  \left\{ \frac{{-2}}{\sqrt{3}\epsilon}  +(\frac{3}{2}-
  \frac\alpha2) + \frac{3}{2} \epsilon + (\gamma+\frac\delta4+\frac32) \epsilon^2
\right\}. 
\end{equation}

The $\mathcal{O}(1/\epsilon)$ term enforces the $|Q| = \pm 1$ constraint; it is due to
the interactions between the charges on the same vertex. At next order,
the honeycomb Madelung energy, with $\alpha \approx 1.5422197$ drives the K1
$\to$ K2 Ising transition. This term reflects the long-range $1/r$ interaction
between the charges (monopoles) 
of different vertices. The $\mathcal{O}(\epsilon^1)$ term does not distinguish
between dimer configurations and hence does not lift their degeneracy; this
instead happens at $\mathcal{O}(\epsilon^2)$, where the dipole-dipole term, with
$\gamma\approx-2.226947$, rather than the monopole-quadrupole term with
$\delta\approx-0.5829489$, leads to a lifting of the degeneracy between dimer
states. (In passing, we note that these constants can be accurately evaluated
using finite-size scaling converging as $1/L^5$ by arranging the unit cell of
the summation to have vanishing monopole, dipole and quadrupole moments).  

The order of this term is easily understood. The charges $\pm q$ are displaced
a distance $\sim \epsilon a$ from the hexagonal vertex they belong to, which
generates a dipole moment of size $|\vec{p}| \sim \epsilon aq$ and hence an
interaction energy of $\mathcal{O}(\epsilon^2)$. This dipole moment points in one of
three directions, namely along the dimer emanating from the site, so that sites
connected by a dimer have aligned dipole moments.  The loop pattern optimises the
local dipolar energy, in that each dimer is oriented favourably with respect to
its four nearest neighbours and due to the convergent nature of
the $1/r^3$ dipolar interactions in d = 2, there is little scope for this state
being destabilised. 

\noindent {\bf Numerics:}~Similarly, in numerics, the regimes are well resolved
for small $\epsilon$, 
where they correspond to extended plateaux in thermodynamic quantities. The
data shown here are for $\epsilon = 0.05$ from Monte Carlo simulations on a
lattice of linear size L = 24, with the interaction summed over periodic copies
up to distance $\Lambda$ = 500 L. We employ a mixture of single-spin flips and the
short loop algorithm,\cite{BarkemaNewman} which seeks a chain of aligned spins
and defines a loop when this chain first crosses itself.
As the presence of different order 
parameters nonetheless renders a fully ergodic algorithm difficult to achieve we 
have pinned down the final ordering transitions by heating and cooling runs, the 
latter at times leading to the formation of large-scale domains.

We find a broad paramagnet $\to$ K1 crossover, followed by a continuous and
hysteresis-free onset of the Ising charge order (K1 $\to$ K2).  These phases
exhibit the predicted entropy, which drops to zero at $T = T_{\rm
  order}$, below which we find the loop crystal, the six-fold degeneracy of which is
due to the (three-fold) translational symmetry breaking, on top of the two-fold
Ising spin reversal symmetry. 

Interestingly, the orientation of the dipoles for a given
translation of the crystal is determined in the transition at
the higher temperature $T_\text{ice}^{II}$. As this
ultimately determines the orientation of the loops, 
one could also refer to it as a chiral
transition. Further, the acceptance rate of the loop moves of our
Monte-Carlo simulations approaches unity in the kagom\'e ice K2 regime,
indicating that it can be regarded as a loop gas. 

\begin{figure}[ttbb]
  \begin{center}
    \includegraphics[width=0.95\columnwidth]{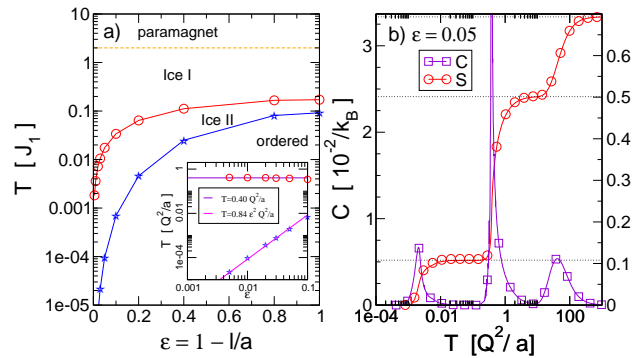}
    \caption{(color online) Numerical results for the dipolar kagom\'e lattice: a) 
      phase-diagram locating the phases (from the peaks in the heat capacity)
      as a function of the 
      length of dipolar magnetic islands $\epsilon=1-l/a$ and temperature
      in units of the NN-interaction $J_1$. 
      The inset details the behaviour for small $\epsilon$: 
      quadratic scaling of the ordering temperature as $T_\text{order}\sim
      \epsilon^2\,q^2/a$ as predicted by Eq.~(\ref{eq:LoopEnergy}), and
      a constant transition temperature $T_\text{ice}^{II} \sim  q^2/a$.
      b) An exemplary trace of the heat-capacity $C$ and entropy per spin $S$
      as a function of temperature for $\epsilon=0.05$. The transitions are
      observed as well separated peaks in the heat capacity. The entropy shows
      broad plateaux matching the theoretical predictions (dotted lines) for
      the different phases.
      }
    \label{fig:simulationResults}
  \end{center}
\end{figure}

In Fig.~\ref{fig:simulationResults}, we show that---at small $\epsilon$---the
temperature scales develop as predicted from the multipole expansions. However,
we note that the two-peak structure of the low-temperature
specific heat persists all the way to the
limit $\epsilon \to 1$ of short dipolar needles. This in particular implies
that the ground state of point like Ising dipoles on the kagom\'e lattice with
in-plane easy axes is given by the loop state. 

\noindent {\bf Magnetic monopoles in artificial spin ice:}~Given the efficient
description provided by the monopoles for the phase diagram 
of the kagom\'e array, it is natural to ask whether the same can be achieved for
artificial (square) spin ice. Indeed, the first salient observation is that the
ground state(s) of that array have zero charge on each vertex---this is
possible as the local coordination is even, whereas it is odd for the kagom\'e
array. This makes contact with Wang's work, whose vertex types I and II
correspond to $Q = 0$ configurations. The ground state (``vacuum'') thus
contains no monopoles.\cite{Wang,MollerMoessner}

This leaves open the question whether monopoles nonetheless appear as effective
low-energy degrees of freedom, as they do in the three-dimensional spin ice
materials $\{Dy, H\!o\}_2$Ti$_2$O$_7$. As was pointed out in
Ref.~\onlinecite{Brazilians}, the
short answer is no. Due to the `antiferromagnetic' order in the ground state,
monopoles cannot move independently as their separation leads to the creation
of a costly domain wall: the monopoles remain confined.  

In our previous work,\cite{MollerMoessner} however, we have proposed a
modification of the square lattice geometry consisting of introducing a height
offset between islands pointing along the two different lattice directions. If
this offset, $h$, is chosen so that the energies of all $Q = 0$ vertices become
degenerate---as $\epsilon\to 0$, the endpoints of the islands then form a
tetrahedron, and $h=\epsilon/\sqrt{2}$---the ordering disappears, and the
monopoles become free to move. Indeed, their interaction is given by
\begin{equation}
V_Q(r) = -\frac{\mu_0}{4 \pi} \frac{(2q)^2}{r}\ .
\end{equation}

Even though we have an array in $d=2$, the interaction follows a $d=3$ Coulomb
law, $1/r$, as the field lines are not confined to the plane. However, 
this is not yet quite the full story. In an appropriate low-temperature
setting, there is in addition an entropic interaction---which hence scales with 
$T$---between the monopoles.\cite{kondev,slideice} This interaction
also is of Coulomb form, but it is two-dimensional in nature, as the entropic
interactions due to the fluctuations of the spin background do not know 
about the existence of a transverse third dimension:  
\begin{equation}
V_s(r) \propto T \ln \left( \frac{r}{a} \right) \ .
\end{equation}
The strength of such a confining potential can obviously be tuned by changing
the temperature---at any rate, this potential is only weak and the
monopoles will look essentially free on reasonable lengthscales. 

{\bf In experiment}, this is true only in principle: a slowing-down of the dynamics
makes the attainment of an equilibrium low-$T$ state difficult.\cite{Wang,MollerMoessner} 
Indeed, the equilibration dynamics of such arrays
generally becomes very sluggish when one encounters energy barriers to flipping
individual islands. In the case of the kagom\'e array, already attaining the K2
charge-ordered crystal presents a formidable challenge in this sense. 
Knowledge of the states to be expected and their
order parameters should nonetheless be useful in identifying equilibration
strategies, especially since a magnetic field is a versatile probe of the spin ices
due to their non-collinear easy axes.\cite{iceprbrc}

The scheme for describing dipolar arrays in terms of magnetic monopoles, as
outlined above, obviously needs to be modified when
$a\epsilon$ becomes comparable to the transverse width of the needles, set to
zero at the outset. However, this will have a substantial effect only on the
$\mathcal{O}(1/\epsilon)$ term in the multipole expansion (\ref{eq:LoopEnergy}), 
and it will in particular not qualitatively change the
relative sizes of the terms in the expansion in this limit. 

Finally, note that the systems studied experimentally typically have a linear
extent of about a few hundred islands.  Given the relatively long-range of the 
interaction---especially that a boundary need not be (magnetic) charge 
neutral---this means that even the ground state may contain domain 
structures interesting in their own right.

{\it Note added in proof.} Recent work by Chern et al.\cite{ChernNote} on the kagome array obtains a phase diagram analogous to ours.

{\bf Acknowledgements:} We are very grateful to Ravi Chandra for
supplying an accurate evaluation of Eq.~\ref{eq:LoopEnergy} and a careful reading of the
manuscript, to him, Claudio Castelnovo and Shivaji Sondhi for
collaboration on related work, and to these and Peter Schiffer, Oleg
Tchernyshyov and Hartmut Zabel for useful discussions.
G.M. acknowledges support from Trinity Hall Cambridge, as well as the hospitality of the University of Colorado at Boulder.


\end{document}